# PRESSURE EFFECT ON THE FORMATION KINETICS OF FERROELECTRIC DOMAIN STRUCTURE UNDER FIRST ORDER PHASE TRANSITIONS

Olga Mazur[1,a)], Leonid Stefanovich[1]

[1] Institute for Physics of Mining Processes of the National Academy of Sciences of Ukraine, Symferopolska st., 2a, Dnipro-5, 49600, Ukraine

[a)]**Author to whom correspondence should be addressed:** o.yu.mazur@gmail.com

## ABSTRACT

Within the framework of Landau – Ginzburg theory the kinetics of domain structure formation in ferroelectrics that undergo first order phase transition was investigated under the influence of hydrostatic pressure. It was established that mechanical action increases the tendency of nonequilibrium system to the formation of stable polydomain structure. Numerical analysis showed that the process of domain ordering can proceed both directly and with formation of short-living metastable polydomain phase. It was found that the incubation period of domain structure formation can be observed at the initial stage of relaxation. Using the $KNO_3$ crystal as an example, it was shown that the time of evolution of the system to the thermodynamic equilibrium is inversely proportional to the pressure value. It was established that in KDP crystals near phase transition point ($T_C - T < 10$ K) the hydrostatic pressure induces the destruction of ferroelectric ordering. It was shown that the value of critical pressure $p_{cr}$ increases with the growth of quenching temperature.

## GRAPHICAL ABSTRACT

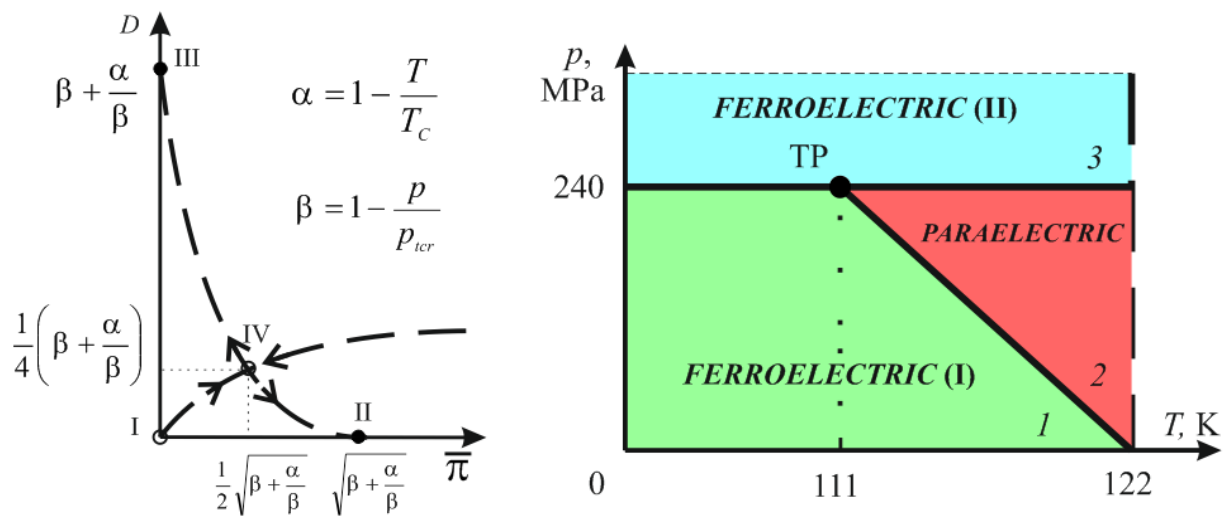

# 1. INTRODUCTION

Ferroelectric materials have a wide application in modern instrumentation. High values of piezo- and pyrocoefficients, large nonlinearity of polarization and minimal dielectric losses allow using them for creation of nonvolatile memory FeRAM, lasers, solar cells and other devices [1, 2].

The unique properties of ferroelectric materials depend on the state of their domain structure, formed under phase transition. Understanding of domain ordering mechanisms opens wide capabilities of ferroelectric samples with specified parameters making. Therefore, many theoretical and experimental works have been devoted to investigation of the processes under ferroelectric phase transition [3–5]. Experiments show that in thin films (up to 200 nm) it is possible to control the processes of domain formation using epitaxial deformation only [6, 7]. Strain engineering techniques are very promising due to the opportunity to improve ferroelectric features of different materials by the changing the type of structural symmetry, shifting of Curie temperature and significant increasing of spontaneous polarization [8]. However, to control the domain structure in thicker examples, a different approach is required, since they cannot be exposed to epitaxial strain. Using the method of rapid cooling in combination with field and mechanical effects at the stage of relaxation of the system to thermodynamic equilibrium can become a promising alternative when working with bulk samples.

Thermodynamic aspects have been studied quite well while the description of phase transition kinetics in strongly nonequilibrium conditions requires a more detailed study. Kinetic approach does not have such universality as thermodynamic describing of equilibrium phase transitions. However, in some cases experimental results cannot obtain sufficient interpretation without studying the process of domain rearrangements under nonequilibrium conditions [4].

In recent years, much more attention has been paid to the experimental study of the kinetics of the ferroelectric phase transition. This is due to the advent of modern equipment that allows us to observe the "time-extended" phase transition. The evolution of the rapidly quenched domain structure of triglycine sulfate (TGS) near the Curie temperature $T_C$ was experimentally studied [5, 9–11]. Numerous theoretical results describing kinetics of phase transition found good experimental confirmation. At the same time previously observed in practice effects received a theoretical justification. Thus, the conclusion established in experiment in [9], that the polarization square is proportional to the quenching depth of the sample $P^2 \sim (T_C-T)$, found theoretical confirmation in [12]. Dependence of the relaxation process on the rate and depth of quenching and the tendency for the enlargement of the domain structure [5], found qualitative theoretical explanation within the framework of kinetic approach developed by us in [13]. In [4] and [13] authors independently of each other showed that the average size of domain region increases according to the square root law.

The next step of research in this area is the study of the relaxation of a rapidly quenched ferroelectric system near the $T_C$ under external influences. Achieving strong nonequilibrium by rapid quenching of the sample from the paraelectric phase to the ferroelectric one has its positive and negative aspects. On the one hand, the nonequilibrium sample becomes very sensitive to even very weak external influences. But on the other hand, it becomes quite brittle. To minimize the risk of destruction, the impact of a very weak stationary electric field on the sample was studied in early investigations [14]. However, the latest experiments show that kinetics of phase transitions can be investigated under the influence of very strong field and pressure [15].

The enlargement of domains in TGS, oriented along the weak electric field superimposed on the sample at relaxation stage under deep quenching, was observed in [14]. These experiments showed that external influences can be applicable to domain structure control. It was shown that in KDP crystals different phase states arise under high pressure (from 2 GPa up to 73 GPa) [15]. In ammonium hydrogen sulfate $NH_4HSO_4$ the second order phase transition turns into the first order phase transition under the influence of pressure $p > 60$ MPa [16]. There was made a direct analogue of kinetic effects observed under thermal transformations with the processes that proceed under pressure in the vicinity of $T_C$ [17, 18].

A detailed review of the study of the effect of high pressures on phase transitions in various ferroelectrics is presented in [19] and [20]. The mechanism of the shift of the Curie temperature under pressure is described in the framework of the thermodynamic approach $T_C$ [19, 20]. The study of the ordering kinetics of ferroelectrics that undergo first order phase transitions under pressure is of a particular interest. There is some tricritical pressure $p_{tcr}$ in such crystals, under which the first order phase transition becomes a second order phase transition and other way round [19–21].

For ferroelectrics that undergo the first order phase transitions the study of kinetic phenomena is associated with mathematical complications compare to those that undergo second order phase transitions [18, 22]. However, this research will allow us to establish the main regularities inherent to ferroelectrics with the first order phase transitions. It is expected that the theoretical results obtained will be able to find practical application in the future.

## 2. FORMULATION OF THE PROBLEM

The fundamental point of our consideration is the study of the relaxation of a ferroelectric sample after its rapid quenching from the high-temperature region (paraelectric phase) to the low-temperature region (ferroelectric phase). As a result of such quenching, the crystal is in a state far from thermodynamic equilibrium. The aspiration of the system to lower its energy makes it pass from a disordered state to an ordered one, which is accompanied by the appearance of small spatial regions containing a ferroelectric phase. As a result, small regions with nonzero polarization $\pm P_z$ spontaneously emerge in the sample at the early stage of ordering when the short-range order relaxation is completely finished. These areas are called 180-degree domains.

Under first order phase transitions the spatial size of domain is usually much larger than the lattice parameter. These small ordered new phase nuclei are nonequilibrium. Therefore, the character of evolution of domains to the thermodynamic equilibrium depends on the initial prehistory of the sample and conditions of its further relaxation. High sensitivity of nonequilibrium system (quenched ferroelectric) to the influence of very weak external effects turns out to be a very useful property in practice. We can change the character of ordering and to obtain the necessary domain structure by varying one or more control parameters (quenching temperature, electric field, mechanical stress) at relaxation stage.

Earlier it was shown that, under first-order phase transitions at an early stage of ordering, sufficiently large domains are formed under the influence of even a very weak electric field [22]. It was established that in such conditions the domain structure is inclined predominantly to the monodomain type of ordering. Formation of polydomain structures both stable and metastable requires the creation of special conditions [22].

In this work we consider the situation when relaxation of quenched sample occurs under the influence of mechanical stresses in the absence of external electric field. We use hydrostatic pressure as the main external influence. The main advantage in favor of the choice of hydrostatic pressure is the possibility of varying its values over a wide range, both in theoretical study and under the conditions of a real experiment. In comparison, the use of uniaxial or tangential mechanical stresses is limited due to the high probability of sample destruction. Hydrostatic pressure does not destroy the crystal, but also transforms it into a denser structure [20].

Let us study how the quenching depth of the sample and the hydrostatic pressure applied at the relaxation stage affect the character of the domain structure evolution. Let us find out what type of ordering the nonequilibrium system will tend to under such mechanical influence.

## 3. GENERAL DESCRIPTION OF THE MODEL

Theoretical description of the kinetics of domain structure significantly depends on the character of phase transition in ferroelectrics considered. Far from $T_C$ both order-disorder (Rochelle salt, TGS, KDP) and soft mode (perovskites) scenarios can be realized. But latest experiments show that near $T_C$ the dynamics of the domain structure is described by the order-disorder regularities independently to the crystal type [23, 24].

Microscopic theories are widely used to describe structural phase transitions in ferroelectric of the order-disorder type [25–27]. The use of general Hamiltonians in these models gives a good description of the features of thermodynamics and phase diagrams, but does not at all touch upon the kinetic aspects of the ordering of a nonequilibrium system. The presence of sufficiently strong short-range correlations in ferroelectrics of the order-disorder type causes a significant dependence of their behavior of the system on the prehistory of the sample and its susceptibility to external influences. Therefore, the use of microscopic approach is accompanied with a number of difficulties associated with the need to refine the data on the nature of crystals: detailing microscopic Hamiltonians, describing possible nonlinear interactions with lattice distortions, clarifying the structures of intermediate phases, as well as the nature of ordering groups in complex crystals [28]. Therefore, for a qualitative description of the kinetics of ordering of the domain structure in the region of the ferroelectric phase transition, it is simpler and preferable to further develop the Landau phenomenological theory of phase transitions [16, 29–32]. The reliability of this approach is confirmed by the fact that the Landau macroscopic model completely predicts the same behavior of the free energy near critical point $T_C$ as the microscopic model [33].

For a theoretical description of the changes occurring in the system during a phase transition, one or several values are introduced, called order parameters $\eta_i$ [34]. In an easiest case of ferroelectric phase transition a projection of vector of spontaneous polarization on the one of crystallographic direction can be chosen as order parameter ($\eta_i = \pm P_z$). There are many models with multicomponent order parameter [35, 36], which give good quantitative description of the dynamics of the domain structure. But the complexity of the considered thermodynamic potentials does not allow obtaining general analytical regularities of the kinetics of ordering of a nonequilibrium system. Therefore, in this work, we consider the case of the formation of the simplest 180° domain structure, which can be described within the framework of the one-component model [37, 38].

For ferroelectrics that undergo first order phase transition close to second order, the nonequilibrium thermodynamic potential in easiest case has a form

$$\Phi\{P_z, \nabla P_z\} = \Phi_0(T) + \int \left[ \varphi(P_z) + \frac{1}{2} G(\nabla P_z)^2 \right] dV. \qquad (1)$$

Here $\Phi_0(T)$ is the part of the thermodynamic potential (1) that does not depend on the order parameter $P_z$, which is the projection of polarization vector $\boldsymbol{P}$ along polar axis $z$. Here and in what follows, we measure the absolute temperature in energy units, assuming the Boltzmann constant equal to unity ($k_B = 1$). Gradient term $G(\nabla P_z)^2$ describes in the first approximation the positive contribution of the energy of domain walls to the thermodynamic potential. The constant $G$ in functional (1) can be estimated as $G \sim T_C^{(0)} R_0^2$, where $R_0$ is the radius of interatomic interaction equal to several lattice constants; $T_C^{(0)}$ is ferroelectric Curie temperature in the absence of pressure $p = 0$.

The density of the thermodynamic potential $\varphi(P_z)$ in functional (1) has the form

$$\varphi(p,T;P_z) = T_C^{(0)} \left[ \frac{1}{2}\alpha(p,T)P_z^2 - \frac{1}{4}\beta(p)P_z^4 + P_z^6 \right], \tag{2}$$

$$\alpha(p,T) = \left(T_c(p) - T\right)/T_C^{(0)}, \quad \beta(p) = (1 - p/p_{tcr}). \tag{3}$$

Here $p$ is hydrostatic pressure, $p_{tcr}$ is the tricritical pressure, quenching function $\alpha(p,T)$ and pressure function $\beta(p)$ are control parameters, which can be used to control the process of domain ordering. Quenching parameter $\alpha(p,T)$ is dimensionless function that describes the closeness of temperature $T$, to which the sample was cooled, to the temperature of ordering $T_C(p)$, which depends on the pressure and is defined as

$$T_C(p) = T_C^{(0)}\left(1 + \frac{p}{\tilde{p}}\right). \tag{4}$$

Here $\tilde{p} = T_C^{(0)}/\gamma$ is some characteristic intrinsic pressure of the crystal, which is related to the coefficient of volumetric compressibility of the material $\kappa = -(\partial V/\partial p)/V$. Parameter $\gamma$ is the baric coefficient, that characterizes the shift of the temperature of ordering $T_C(p)$ under the influence of hydrostatic pressure $p$, and is defined as

$$\gamma = \left.\frac{\partial T_C}{\partial p}\right|_{p=0} \sim \frac{T_C^{(0)} \kappa r_0}{\varepsilon_{at}}. \tag{5}$$

Here $\varepsilon_{at}$ is the energy of interatomic interaction [18]. Pressure $p$ can both increase and decrease the temperature of ordering $T_C(p)$ depending on the relation between values and signs of elastic coefficients of the crystal. When $\gamma > 0$ the hydrostatic pressure shifts the Curie temperature to higher temperatures and when $\gamma < 0$ – to the region of lower temperatures. Too high pressure can damage the crystal. Therefore we assume that the value of external pressure $p$ is much smaller than the intrinsic pressure $p \ll \tilde{p}$.

Parameter $\beta(p)$ in (2, 3) is dimensionless function that describes the vicinity of the pressure imposed on ferroelectric $p$ to the tricritical value $p_{tcr}$ [19–21]. The negative sign of the coefficient $\beta(p)$ in (2) characterizes the first-order phase transition. At pressures close to the tricritical value $p \to p_{tcr}$, coefficient $b \to 0$ tends to zero. At $p = p_{tcr}$ parameter $b$ vanishes. With a further increase in pressure, the $b$ term changes sign, and the first-order phase transition becomes a second-order phase transition, i.e. changes its character. Therefore, we will consider pressures less than tricritical value $p < p_{tcr}$ to describe the kinetics of ordering of domain structure under first order phase transition.

The relaxation process of such a system under isothermal conditions taking into account the imposed pressure can be described using the Landau – Khalatnikov equation [18, 22]. This relaxation equation allows us to obtain the evolution equation for the order parameter $\pi$

$$\frac{\partial \pi}{\partial \tau} = \Delta \pi + \alpha(p,T)\pi + \beta(p)\pi^3 - \pi^5. \tag{6}$$

Here $\pi = P_z/P_s$ is a dimensionless order parameter ($P_s$ – saturation polarization on the hysteresis curve), $\Delta$ – Laplace operator. Here $\tau = t/t^*$ is dimensionless time, where $t^*$ is the characteristic time of the system rearrangement (for example, the displacement of an atom). Random character of distribution of initial inhomogeneities of polarization formed by rapid quenching requires to add the initial condition $\pi(r,\tau)|_{\tau=o} = \pi(r,0) = \pi(r)$ to the equation (6), where $\pi(r)$ is the random function of coordinates. Consequently, equation (6) describes the time evolution of random field of order parameter.

Statistical approach can be used to describe the kinetics of ordering of the quenched system due to the random spatial distribution of order parameter over the sample. For this we have to find the average value of order parameter $\langle \pi(r,\tau) \rangle \equiv \bar{\pi}(\tau)$ and two-point correlation function $K(s, \tau)$, where $s = r'–r$. Angle brackets here denote the averaging process over the ensemble of realizations of the random field of the order parameter. The decoupling procedure of correlators of fourth order to correlators of second order developed for nonequilibrium system was described in detail in [22]. However, it can be applicable to the case when the pressure is imposed on the system ($p \neq 0$) due to its universal character. As a result we obtain the closed system of equations for functions $\bar{\pi}(\tau)$ и $K(s, \tau)$:

$$\begin{cases} \dfrac{d\bar{\pi}}{d\tau} = \dfrac{1}{2}\left[\alpha(p,T)\bar{\pi} + 3\beta(p)\bar{\pi}K(\tau) - 5\bar{\pi}K^2(\tau) + \beta(p)\bar{\pi}^3 - 10\bar{\pi}^3 K(\tau) - \bar{\pi}^5\right] \\ \dfrac{\partial K(s,\tau)}{\partial \tau} = \Delta K(s,\tau) + \left[\alpha(p,T) + \beta(p)(3\bar{\pi}^2 + K(\tau)) - 5\bar{\pi}^2(2K(\tau) + \bar{\pi}^2) - K^2(\tau)\right]K(s,\tau) \end{cases} \quad (7)$$

System of equations (5) contains two control parameters to direct the ordering process in practice: hydrostatic pressure $p$ and quenching parameter $\alpha$. It cannot be solved analytically because of nonlinear character of right sides of this system. But with the help of the Fourier transform of the second equation, it is reduced to a system of ordinary differential equations of the following form

$$\begin{cases} \dfrac{d\bar{\pi}}{d\tau} = \dfrac{1}{2}\left[\alpha(p,T)\bar{\pi} + 3\beta(p)\bar{\pi}D(\tau) - 5\bar{\pi}D^2(\tau) + \beta(p)\bar{\pi}^3 - 10\bar{\pi}^3 D(\tau) - \bar{\pi}^5\right] \\ \dfrac{dD(\tau)}{d\tau} = \left[\alpha_{eff}(p,T,\tau) + 3\beta(p)\bar{\pi}^2 + \beta(p)D(\tau) - 10\bar{\pi}^2 D(\tau) - 5\bar{\pi}^4 - D^2(\tau)\right]D(\tau) \end{cases} \quad (8)$$

Here $D = D(\tau) = K(0,\tau)$ – dispersion of order parameter. We showed that the character of formed ferroelectric domain structure also depends on average size of initial inhomogeneities of order parameter, formed at quenching stage [12]. Evolution character of correlation radius $\rho(\tau)$ (average characteristic scale of ordered regions dimensionless on $R_0$) defines by time dependence of coefficient $\alpha_{eff}(p,T,\tau)$:

$$\alpha_{eff}(p,T,\tau) = \alpha(p,T) - \rho(\tau), \qquad \rho(\tau) = \left(2\tau/3 + \rho_c^2(0)\right)^{-1}. \quad (9)$$

Here $\rho_c(0)$ is the initial correlation radius (average size of domain nuclei) [12, 13].

## 4. ASYMPTOTIC BEHAVIOUR OF THE SYSTEM AT LARGE TIMES

Ferroelectric properties are most clearly manifested near the phase transition temperature. In this case the initial correlation length in dimensionless units is small, i.e. $\rho_c^2(0) << 1/\alpha << d^2$, where $d$ is characteristic size of crystallite. The initial correlation radius $\rho_c(0)$ and characteristic size of crystallite are dimensioned for the interatomic interaction radius $R_0$.

The nonlinearity of the resulting system of differential equations (8) provides a variety of scenarios for the evolution of the domain structure. It allows us to trace all the stages of its formation. Due to the complexity of the system (8) it can be solved by numerical methods only. But it is a very useful tool for a detailed study of the kinetics of ordering of the domain structure of each certain ferroelectric. Numerical solvation of system (8) allows us to predict the evolution scenario of domain ordering depending on different relaxation conditions and pressure value in particular.

The analytical solving and establishing of general kinetic aspects of the domain ordering process can only be done using the asymptotic approach. Therefore, the simplified system of equations was studied at large times, when the function $\alpha_{eff}(p,T,\tau)$ ceases to depend on time and become equal to $\alpha_{eff}(p,T,\tau) = \alpha(p,T)$. At large times, i.e. when $\tau \to \infty$, left sides of equations in (8) go to zero: $(d\bar{\pi}/d\tau) \to 0$, $(dD/d\tau) \to 0$. As a result, we obtain the nonlinear algebraic system of equations

$$\begin{cases} \bar{\pi}\left(\alpha(T,p)+3\beta(p)D-5D^2\right)+\bar{\pi}^3\left(\beta(p)-10D\right)-\bar{\pi}^5 = 0 \\ \left(\alpha(T,p)+3\beta(p)\bar{\pi}^2-5\bar{\pi}^4+\beta(p)D-10\bar{\pi}^2D-D^2\right)D = 0 \end{cases}. \quad (10)$$

Solutions of the system of equations (10) give us the coordinates of singular (stationary) points in a plane ($\bar{\pi}$, $D$):

$$\begin{cases} (\text{I}): \ \bar{\pi}=0; \ D=0 \\ (\text{II}): \ \bar{\pi}=\left(\beta+\dfrac{\alpha}{\beta}\right)^{1/2}; \ D=0 \\ (\text{III}): \ \bar{\pi}=0; \ D=\beta+\dfrac{\alpha}{\beta} \\ (\text{IV}): \ \bar{\pi}=\dfrac{1}{2}\left(\beta+\dfrac{\alpha}{\beta}\right)^{1/2}; \ D=\dfrac{1}{4}\left(\beta+\dfrac{\alpha}{\beta}\right) \end{cases} \quad (11)$$

For $\alpha < 0$, i.e. in high-temperature region $T > T_C$ the system is in the disordered state corresponding to the paraelectric phase. In the phase pattern, there is only one singular point with coordinates ($\bar{\pi} = 0$, $D = 0$), which is a stable node.

After rapid quenching to the temperature region $T < T_C$, i.e. for $\alpha > 0$ the system turns out at nonequilibrium state and becomes unstable regarding to the transition into ferroelectric phase. The nonlinear character of the system (10) provides the whole complex of singular points (Fig. 1(a)).

Singular point I becomes an unstable node. Immediately after quenching, the nonequilibrium system, as a rule, turns out to be in the vicinity of this point, from where it begins its relaxation to the state of thermodynamic equilibrium (Fig. 1).

Singular point II is a stable node (Fig. 1(a)). This point characterizes a thermodynamically stable monodomain structure.

Singular point III is a stable node too (Fig. 1(a)). However, this point describes a thermodynamically stable polydomain structure with zero average polarization $\bar{\pi}$. The degree of inhomogeneity of the formed domain structure defines by the quenching depth and the value of hydrostatic pressure imposed on the system.

Singular point IV is a saddle point (Fig. 1(a)). It corresponds to the unstable short-living polydomain state with a pronounced asymmetry of volume fractions of domains of opposite signs. The system continues to relax to the one of the thermodynamically stable states after a short slowing down the speed of its movement in the vicinity of saddle point IV.

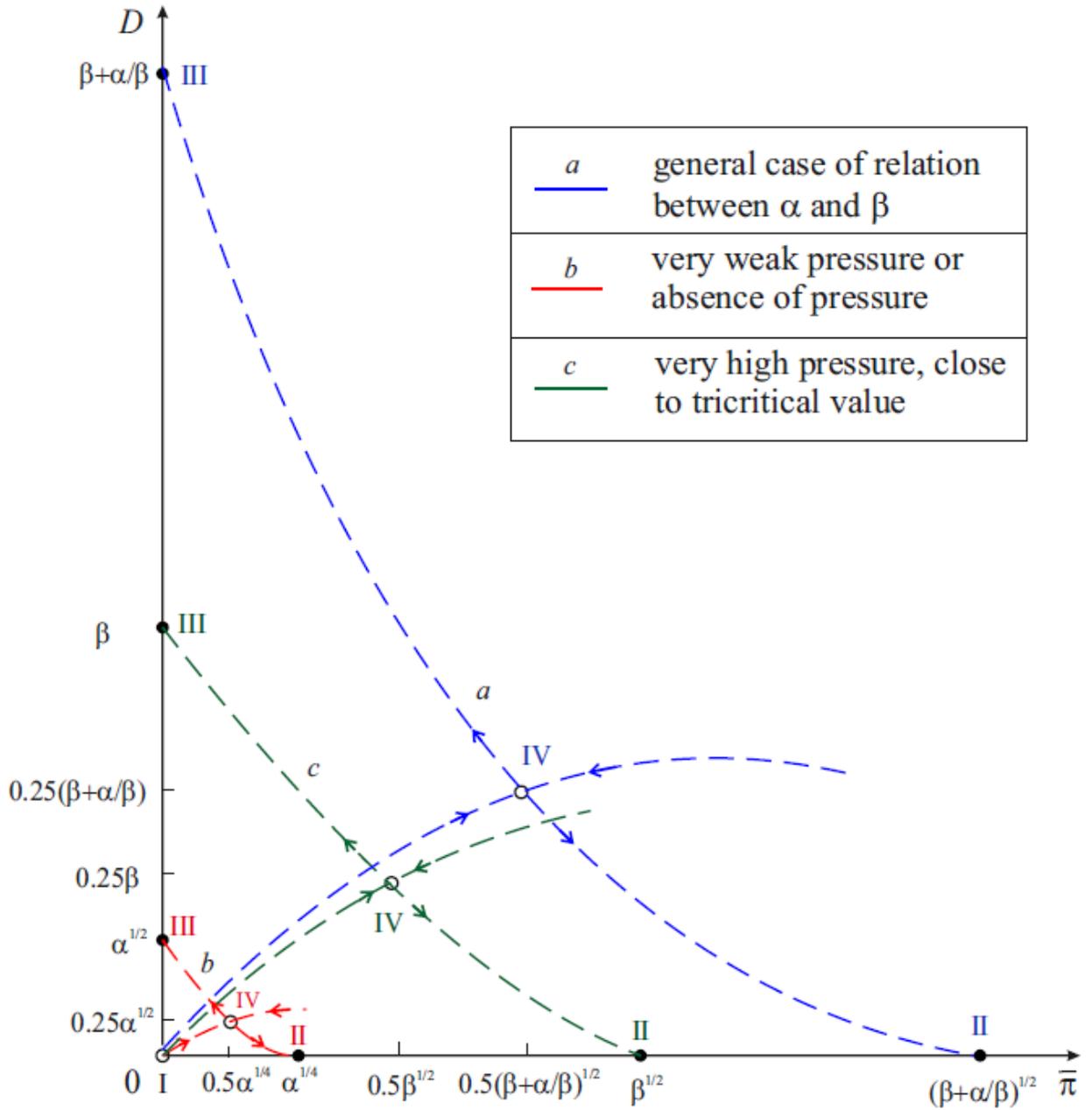

**Fig.1.** Qualitative phase pattern of the system in variables $(\bar{\pi}, D)$ under different relation between parameters α and β. Singular points of the system are indicated by Roman numerals. The dashed lines indicate the separatrices. (*a*) - general view of the phase pattern for arbitrary values of the parameters α and β; (*b*) - phase pattern under the influence of very weak hydrostatic pressure or in its absence; (*c*) - the effect of strong hydrostatic pressure close to the tricritical value.

Within the framework of this problem, the situation is considered when the system relaxes under the conditions of the hydrostatic pressure imposed on the sample. However, in order to trigger the mechanism of nucleation of microregions oriented along the field, it is assumed that at the stage of quenching (prehistory) the system was under the influence of a weak external electric field $\varepsilon_q$. The "quenching" electric field $\varepsilon_q$ forms domains with corresponding sign of polarization vector during the quenching of the sample. As a result, at the initial stage of relaxation the average polarization of the system becomes nonzero $(\bar{\pi}_0 \neq 0)$. In this problem it is assumed that polarization vector of this "quenching" field $\varepsilon_q$ is directed along polar axis of the crystal $(\bar{\pi}_0 > 0)$. Therefore, we observe the relaxation of the sample in the first quadrant of phase pattern (Fig. 1). If "quenching" field $\varepsilon_q$ is directed opposite to the polar axis of the crystal, polarization vector has a corresponding sign $(\bar{\pi}_0 < 0)$ and the evolution of domain structure occurs in the second quadrant of phase pattern.

Already at the stage of a qualitative analysis of the system of equations (10), it is possible to reveal some distinctive regularities of the process of ordering the domain structure of ferroelectrics that undergo first-order phase transitions. These ferroelectrics are inclined mainly to the formation of stable monodomain structure [22]. However, under the hydrostatic pressure the probability of formation of polydomain structure both stable and metastable increases.

The evolution character of domain structure depends on prehistory of the sample at quenching stage. However, the thermal and mechanical effects provide the greatest influence on the ordering process at relaxation stage of the system to the state of thermodynamic equilibrium.

The location of singular points in phase pattern (Fig. 1) significantly depends on relation between parameters α and β. Therefore, for a more detailed study of the main scenarios of the evolution of the system to the state of thermodynamic equilibrium, several possible conditions should be considered.

### 4.1. Effect of a very weak hydrostatic pressure $p \ll p_{tcr}$, i.e. $p \rightarrow 0$

In the framework of considered problem the quenching depth cannot be too strong. Therefore, this limiting case corresponds to the situation, when the system relaxes under the influence of very weak pressure both in the vicinity of $T_C$ ($\alpha > \beta$) and under the deep quenching ($\alpha \gg \beta$). In this case singular points of the system (10) has a form

$$\begin{cases} (I): \ \bar{\pi} = 0; \ D = 0 \\ (II): \ \bar{\pi} = \alpha^{1/4}; \ D = 0 \\ (III): \ \bar{\pi} = 0; \ D = \alpha^{1/2} \\ (IV): \ \bar{\pi} = \frac{1}{2}\alpha^{1/4}; \ D = \frac{1}{4}\alpha^{1/2} \end{cases} \quad . \tag{12}$$

The location of singular points defines by the quenching parameter α and almost does not depend on the value of hydrostatic pressure $p$. Formation of both mono- and polydomain structures is equiprobably. Due to the low dispersion, the formed polydomain structure corresponds to large domains (Fig. 1(b)). Indeed, the pressure imposed on the sample at relaxation stage is so small, that it does not disturb the reversal of dipoles and overgrowth of domain regions with one or the other polarization vector.

### 4.2. Effect of relatively strong pressure ($0 \ll p < p_{tcr}$)

It is assumed that pressure $p$ does not destroy the ferroelectric ordering. In this case the evolution of domain structure depends on both quenching temperature and value of hydrostatic pressure $p$ imposed on the sample. Therefore, singular points of the system of equations (10) are defined by coordinates (11). The phase pattern expands with the value of hydrostatic pressure (Fig. 1(a)). Significant growth of dispersion of singular point III is observed (Fig. 1(a)). This means that under such conditions the hydrostatic pressure grinds the ordered regions formed at the quenching stage, promoting the formation of stable polydomain structure. The effect of the quenching temperature becomes less and less noticeable with increasing pressure.

### 4.3. Effect of high pressure, close to tricritical value ($p \sim p_{tcr}$)

High hydrostatic pressure is capable repolarize the sample both in the vicinity of $T_C$ ($\beta \gg \alpha$) and under deeper quenching ($\beta > \alpha$). Therefore, the location of singular points almost does not depend on the value of parameter $\alpha$ and defines as

$$\begin{cases} (I): \bar{\pi} = 0; \quad D = 0 \\ (II): \bar{\pi} = \beta^{1/2}; \quad D = 0 \\ (III): \bar{\pi} = 0; \quad D = \beta \\ (IV): \bar{\pi} = \frac{1}{2}\beta^{1/2}; \quad D = \frac{1}{4}\beta \end{cases} \quad (13)$$

This limiting case (Fig. 1(c)) differs from the general case only by the amendment $\alpha/\beta$ (Fig. 1(a)). Under relatively high pressure this amendment is small and cannot influence on the process of ordering of domain structure. The system will tend to form a predominantly stable polydomain state with small domain sizes. This is in good correspond with the conclusion that at pressures close to the tricritical value, the first-order phase transition acquires the features of a second-order phase transition. In this case formation of stable polydomain structure is preferable compare to monodomain type of ordering [18].

According to the results of the performed qualitative analysis, it can be concluded that in order to achieve a monodomain type of ordering in ferroelectrics that undergo first-order phase transitions, it is better to perform relaxation in the absence of external influences. This is consistent with our earlier results in [22]. However, it is necessary to use hydrostatic pressure to obtain stable polydomain structures in these crystals at the stage of relaxation to the state of thermodynamic equilibrium. An inversely proportional relationship is observed between the size of the domains and the pressure value. However, the strength properties of a particular crystal as well as by quenching conditions constrain the possibility of using of high values of hydrostatic pressure.

### 5. ANALYSIS OF COMPLETE EVOLUTION EQUATIONS

The use of the phase pattern concept made it possible to determine the stationary states of the system in an analytical way and to establish the general laws of the process of ordering of the domain structure. However, in order to trace all stages of relaxation of the system to the state of thermodynamic equilibrium, it is necessary to solve the complete system of evolutionary equations (8). For this, the evolution of a nonequilibrium system under the impact of hydrostatic pressure was considered by numerical methods. The complete system of differential equations (8) was solved in MatLab package by Runge-Kutta method.

The numerical analysis was made using potassium nitrate $KNO_3$ and potassium dihydrogen phosphate $KH_2PO_4$ (KDP) crystals, which have different signs of baric coefficient.

Within the framework of investigation it was assumed that a weak external electric field $\varepsilon_q$ is imposed on the sample at the quenching stage and is turned off after the formation of initial inhomogeneities of the critical size. Thus, both the initial polarization $\bar{\pi}_0$ and initial dispersion $D_0$ are nonzero. At the end of quenching, the relaxation of the system occurs only under influence of all-round (hydrostatic) pressure. Let us trace all the stages of the formation of the domain structure of a ferroelectric and find out how the hydrostatic pressure and the quenching depth affect the relaxation process. Let us compare the features of the ordering of the domain structure of ferroelectrics with different signs of the bulk compressibility coefficient.

The numerical analysis was made for the case, when the inducing initial polarization $\pi_0$ "quenching" electric field $\varepsilon_q$ is directed along the polar axis of the crystal. Therefore, all relaxation processes occur at the first quadrant of the phase pattern (Fig. 1).

### 5.1. Potassium nitrate KNO$_3$

Potassium nitrate crystal KNO$_3$ is a promising material as a ferroelectric component in the creation of non-volatile memory devices. The small switching time (20 ns), low value of switching potential (5 V) and square hysteresis loops are the most important practical features [39].

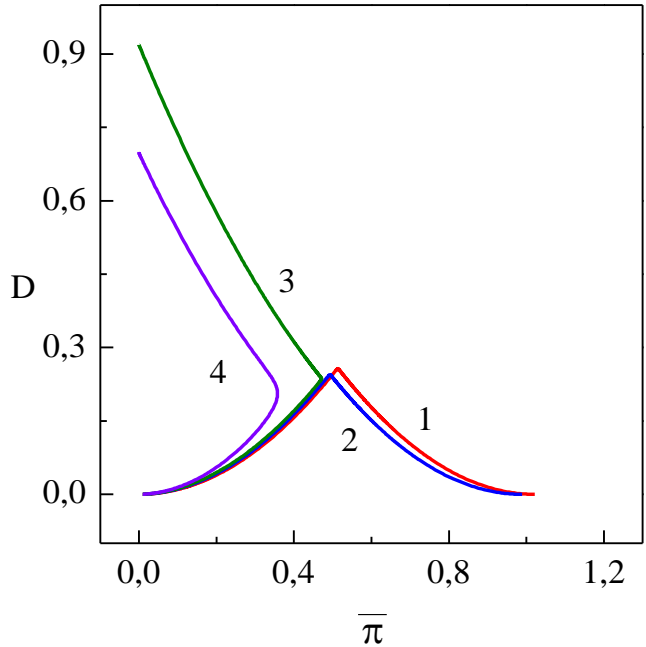

**Fig. 2.** Phase trajectories of the system, which describe the dynamics of the formation of domain structure of KNO$_3$. Quenching temperature $T_1=380$ K. Quenching conditions in dimensionless units: initial average polarization $\bar{\pi}_0 = 0.01$, initial dispersion $D_0 = 0.0002$, initial average size of polarization inhomogeneity $\rho_0 = 3$. Phase trajectories 1 – 4 correspond to the values of pressure 0, 0.1, 0.2, 0.79 GPa respectively

Potassium nitrate crystal undergoes the first order phase transition of order-disorder type at the temperature $T_C = 398$ K [40]. In bulk samples the ferroelectric phase remains stable at temperatures 398 – 378 K. It is interesting that in thin films of potassium nitrate the ferroelectric state remains stable at even wider temperature range 403 – 273 K [39]. The temperature range of ferroelectric stability in a bulk ferroelectric can be expanded by doping. However, in the framework of our problem an ideal crystal in the absence of defects and impurities is considered. Therefore, the numerical analysis was made at corresponding temperature range.

The Curie temperature $T_C(p)$ of potassium nitrate increases with pressure $p$ ($\gamma > 0$). The following physical characteristics of the KNO$_3$ crystal were used during numerical analysis: Curie temperature $T_C = 398$ K [24], Debye temperature $\Theta_D = 270$ K [41], baric coefficient $\partial T_C/\partial p = +220$ K/GPa [19], activation energy $Q = 0.89$ eV [42], tricritical pressure $p_{tcr} = 0.8$ GPa [43].

The numerical analysis showed that formation of domain structure under hydrostatic pressure sufficiently depends on the quenching depth α. When the sample is quenched in the vicinity of Curie temperature ($T_C - T_1 \leq 1$ K) even very weak pressure $p$ promotes the formation of a thermodynamically stable polydomain ordering state, which corresponds to singular point III in the phase pattern (Fig. 1). The higher the hydrostatic pressure, the faster a stable polydomain structure forms in the system.

Stable monodomain structure, corresponding to singular point II in phase pattern (Fig. 1), forms only in the absence of pressure $p = 0$ or when the quenching field $\varepsilon_q$ is strong. It is confirmed by the conclusions in [22].

Thus, under the not deep quenching the hydrostatic pressure leads the relaxation system into polydomain type of ordering at any pressure in the range $0 < p < p_{tcr}$. The magnitude of pressure $p$ influences only on the general view of the trajectory of the system and also on velocity of relaxation to the state of thermodynamic equilibrium.

With increasing of quenching depth, i.e. with an increase in the parameter $\alpha$, the role of hydrostatic pressure decreases, and at different values of it, both polydomain and monodomain structures can form (Fig. 2). With deep quenching, cooling process takes longer, and large initial domains have time to form in the system. These domains cannot be destroyed by weak pressure, which is applied to the sample at the stage of early relaxation. Thus, the scenario of the evolution of the system under the influence of a relatively weak pressure (curve 2 in Fig. 2) is similar to the case when there is no pressure (curve 1 in Fig. 2) - a monodomain structure is formed corresponding to the singular point II in the phase pattern (Fig. 1).

However, even with deep quenching the system remains susceptible to external influences. With increasing pressure $p$, the evolution scenario of the system changes sharply, and as a result, a stable polydomain structure is formed (curve 3 in Fig. 2). When the magnitude of pressure imposed on the sample is close to tricritical value $p_{tcr}$ the smoother and faster transition to polydomain ordering state is observed (curve 4 in Fig. 2). Such behavior of the system can be explained by the fact that under pressures close to tricritical value $p \to p_{tcr}$ the phase transition begins to acquire the features of a second order phase transition, in which polydomain structures are predominantly formed [18].

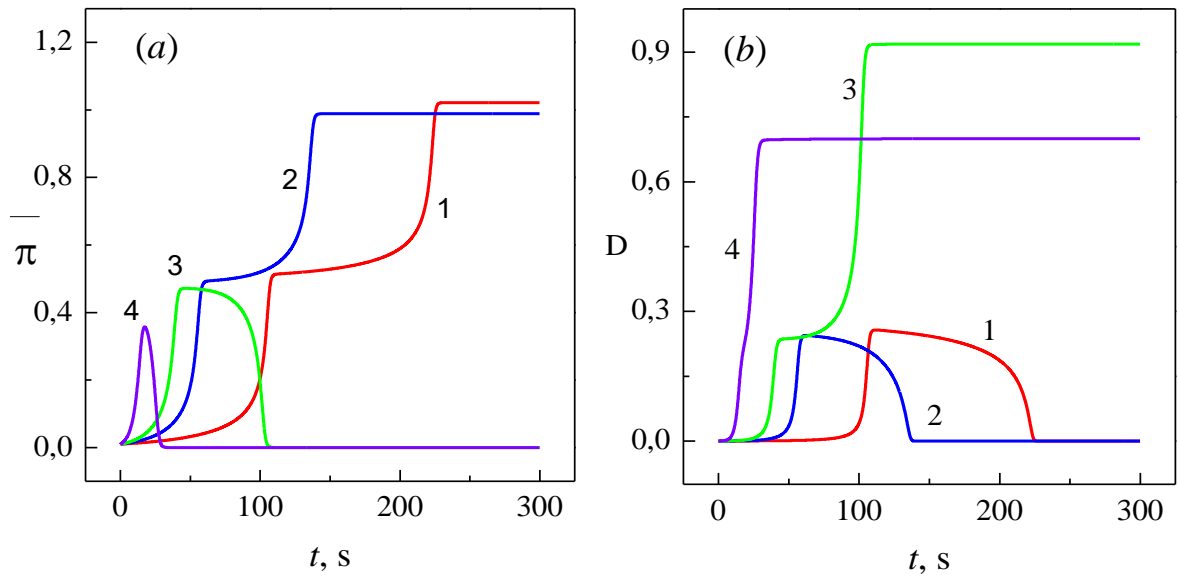

**Fig. 3.** Evolution of average polarization (a) and its dispersion (b) of ferroelectric $KNO_3$ under the same conditions as in Fig.2. Steps (Fig. 3a) and plateaus (Fig. 3b) on evolution curves 1, 2 and 3 characterize the kinetic deceleration of the system near the saddle point and corresponds to the polydomain states with the pronounced asymmetry of domains of the opposite sign of polarization vector. Curve 4 shows the direct evolution of the system into the state of thermodynamic equilibrium without metastable stage near saddle point.

Evolution curves for average polarization $\bar{\pi}_0$ and dispersion $D_0$ showed that under deep quenching the general time of relaxation of the system takes less than five minutes (Fig. 3(a) and 3(b)). It was established that process of domain ordering can proceed nonmonotonically with the formation of short-lived polydomain phases with a pronounced asymmetry of volume fractions of domains with different signs (curves 1,2,3 in Fig. 3(a) and 3(b)). Such metastable states characterize the kinetic deceleration of the system in the vicinity of saddle point IV (Fig. 1). The existence time of the system in this state can be determined by the length of the intermediate step (curves 1,2 in Fig. 3(a) and 3(b)) or plateau (curve 3 in Fig. 3(a) and 3(b)) in evolution curves.

Numerical analysis showed that in ferroelectrics $KNO_3$ the ordering of domain structure rarely occurs monotonically and the system is prone to the formation of asymmetric metastable phases (Fig. 2, 3). But the observation of such states is possible only under some conditions (temperature, pressure, electric field) because the coming of the system in the vicinity of the saddle point IV (Fig. 1) strongly depends on the external influences. Such behavior of the system can be associated with Barkhausen effect – intermittent change of the electrical state of ferroelectric crystal caused by domain nucleation and fast forward growth [44, 45]. In ferroelectrics, the Barkhausen effect can arise both under the action of an electric field and in its absence during a phase transition, when the process of rearrangement of the domain structure inside the ferroelectric phase occurs. Also, jumps of polarization reversal can be caused by the application of mechanical stresses to the sample [44].

The mechanism of the polarization jump during the formation of the domain structure is associated with the fact that there are always regions in a stressed state inside the crystal. These regions are nucleation centers, since when an external pressure or electric field is applied, the reorientation of spontaneous polarization in them will occur more easily than in other regions of the crystal. Growing rapidly, these nuclei also cause jumps in polarization reversal. But such Barkhausen effect can occur only when the growth of domains is energetically favorable.

At a certain moment, the process of formation and growth of new domains reaches its peak (Fig. 2) and the mobility of domain walls decreases until the appearance of regions with linear polarization (curves 1–3 in Fig. 3). The kinetic deceleration of the system near the saddle point occurs by the linearity of both polarization (Fig. 3(a)), and its dispersion (Fig. 3(b)). In the linear region of the polarization and dispersion curves, the domain structure is practically not rearranged, and the ferroelectric behaves like a paraelectric. But such an unbalanced state cannot be energetically beneficial. After the kinetic deceleration of the system, a stage of stabilization of the domain structure is observed, as a result of which one of the thermodynamically stable states is formed.

Not only the scenario of the domain structure evolution, but also its duration (relaxation time $t$) depends on the value of external parameters. Fig. 4 shows how the relaxation time of domain structure depends on the value of the pressure that was imposed on the sample at different quenching depth. When the sample is quenched in the vicinity of $T_C$ (curve 1 in Fig. 4) the relaxation time depends significantly on the pressure. The more pressure is applied to the sample, the faster the domain structure is formed. With the increase in the quenching depth this nonlinearity decreases (curves 1-5 in Fig. 4). When the sample is deeply quenched (curves 4, 5 in Fig. 4) the relaxation time is almost independent of pressure. It is important to admit that nonlinear dependence of relaxation time $t$ from the pressure $p$ occurs only under values $p \leq 0.2$ GPa. When pressure is $p > 0.2$ GPa it practically does not influence on the duration of domain structure formation at any quenching depth (linear region in Fig. 4 for curves 1-5).

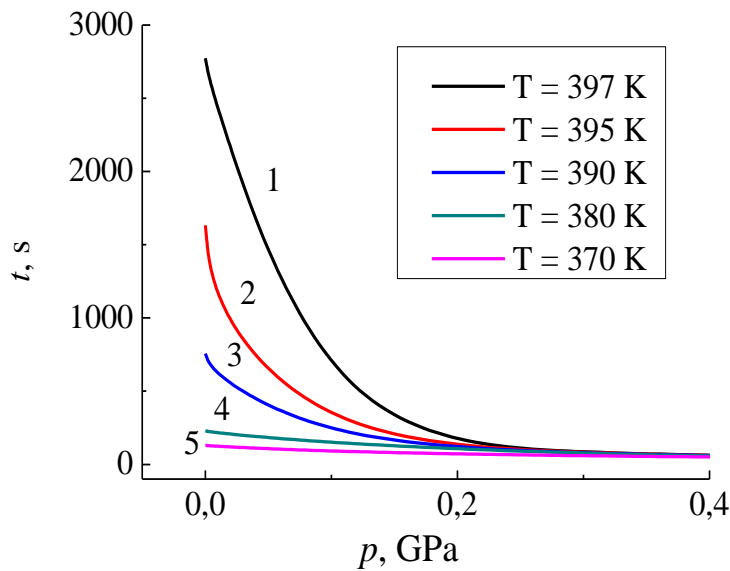

**Fig. 4.** Dependence of relaxation time $t$ of domain structure of $KNO_3$ crystal from the value of hydrostatic pressure $p$ for different quenching depth $T$. Curves 1 - 5 correspond to quenching temperature $T$: 397K; 395K; 390K; 380K; 370K respectively

Thus, hydrostatic pressure, like any other effect, accelerates the process of ordering of domain structure. However, this effect is pronounced only under conditions of not deep quenching and under pressures $p \leq 0.2$ GPa (Fig. 4).

This once again confirms the fact that during shallow quenching the system is very sensitive to small external influences (in this case, to the hydrostatic pressure imposed on it).

### 5.2. Potassium dihydrogen phosphate KH$_2$PO$_4$ (KDP)

Ferroelectrics with a KDP-type structure are very promising materials due to the possibility of growing single crystals of high optical quality from an aqueous solution. Due to the complexity of the crystal structure of KDP, the microscopic description of the dynamics of these ferroelectrics requires the introduction of 48 unit vectors (two formula units per primitive cell) [46]. However, phenomenological consideration within the Landau theory allows us to describe reliably most of the properties of KDP in the vicinity of the phase transition [47].

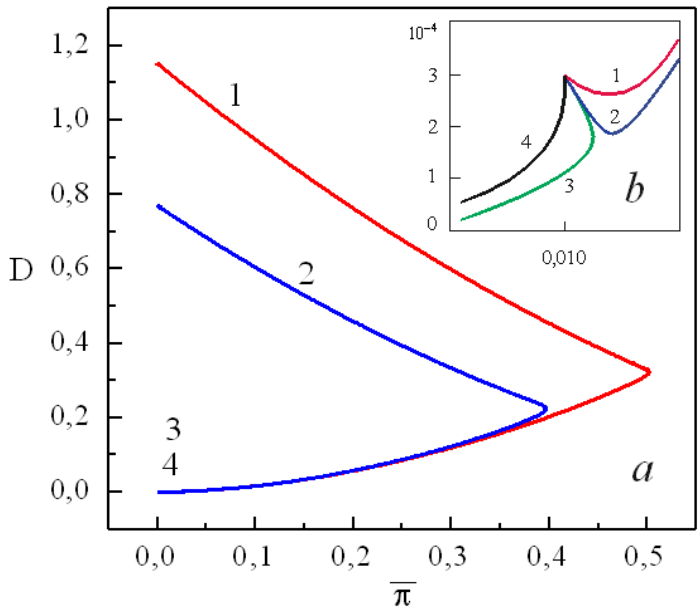

**Fig. 5.** Phase trajectories of the system describing the dynamics of the formation of the domain structure of a ferroelectric KDP: *a* – complete phase pattern; *b* – large-scale image of the initial relaxation phase of the system. Quenching temperature $T_1=121$ K. Quenching conditions in dimensionless units: initial average polarization $\bar{\pi}_0 = 0.01$, initial dispersion $D_0 = 0.0002$, initial size of polarization inhomogeneity $\rho_0 = 3$. Phase trajectories 1 – 4 correspond to the value of pressure 0.001, 0.023, 0.024, 0.03 GPa respectively. Curves 1 and 2 show the appearance of a ferroelectric phase and the evolution of the domain structure to a stable polydomain state. Curves 3 and 4 which are reflected in the sidebar (*b*) characterize the situation when the pressure suppresses the nuclei of ferroelectric domains formed after quenching, and the system returns to the disordered paraelectric phase.

The KDP crystal has found a wide application, both as a model object and in experimental studies of the kinetics of ordering of the domain structure in the phase transition region. It was shown that as a result of quenching in KDP crystals, not just a regular structure, but individual blocks with 180º domains are formed, which gradually grow and can have different shapes relative to each other [4]. The average size of domain block increases according to the square root law, which corresponds with our results in [22].

The Curie temperature $T_C(p)$ of KDP decreases with pressure $p$ ($\gamma < 0$). The following physical characteristics of the KDP crystal were used during numerical analysis: Curie temperature $T_C = 122$ K [46], Debye temperature $\Theta_D = 169$ K [48], baric coefficient $\partial T_C/\partial p = -45$ K/GPa [19], activation energy $Q = 0.22$ eV [48], tricritical pressure $p_{tcr} = 0.24$ GPa [49].

Compare to the potassium nitrate KNO$_3$, in KDP crystal both in the vicinity of $T_C(p)$ and under deep quenching ($T_C - T_1 \leq 20$ K), the domain structure evolves to the polydomain state only. Numerical analysis showed that in these ferroelectrics not a quenching depth α, but exactly the value of pressure $p$ plays a decisive role in the relaxation process. The evolution process can proceed both directly and with the formation of short-living asymmetric metastable phases (curve 1 in Fig. 5).

However, compare to the KNO$_3$, there is the possibility of destruction of ferroelectric phase under pressure in KDP crystals, especially in the vicinity of $T_C(p)$. Numerical analysis showed that there is some critical pressure $p_{cr}$ in the case of not deep quenching ($T_C - T_1 \leq 20$ K). This pressure characterizes the destruction of ferroelectric ordering, but not a damage of crystal. Therefore, the magnitude of critical pressure is much less than the value of characteristic intrinsic pressure of the crystal $p_{cr} << \tilde{p}$.

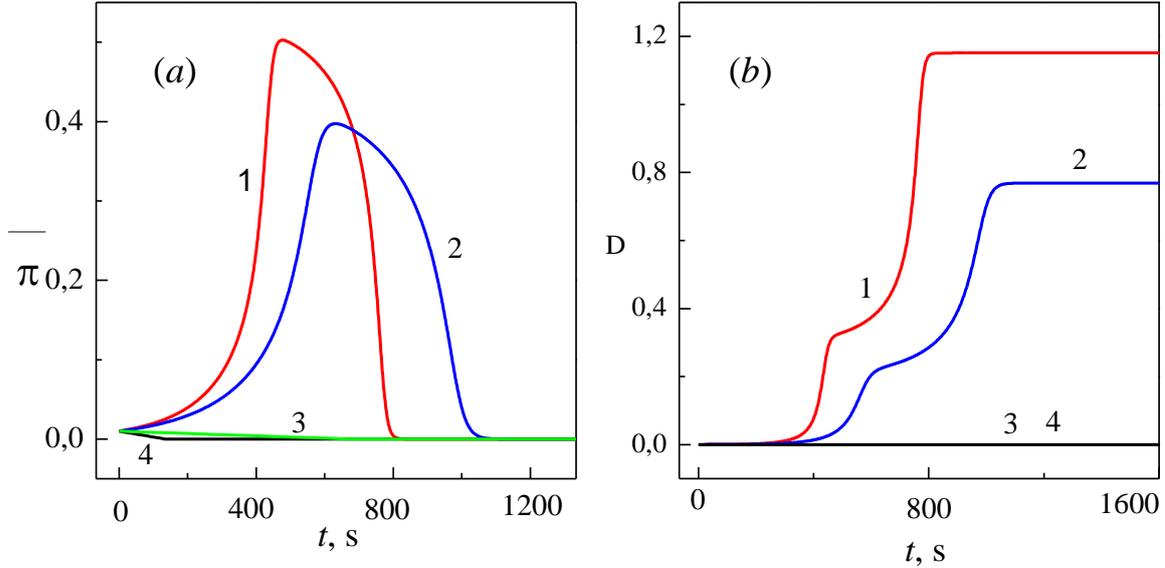

**Fig. 6.** Evolution of average polarization (*a*) and its dispersion (*b*) of ferroelectric KDP under the same conditions as in Fig.5. Curves 1 and 2 show the evolution of the average polarization (*a*) and its dispersion (*b*) in the ferroelectric phase. Curves 3 and 4 correspond to the transition of the system to a disordered paraelectric state immediately after quenching.

Phase pattern shows (Fig. 5) that under very weak pressure ($p = 1$ MPa) the system gradually evolves to the stable polydomain 1 in Fig. 5). However, a tendency to disorder appears under larger pressure ($p = 23$–24 MPa). At $p = 3$ MPa the internal state (curve resistance of the crystal turns out to be sufficient so that the nuclei of domains formed during quenching are not destroyed, and the domain structure continues to develop (curve 2 in Fig. 5). Under pressures higher than critical value $p > p_{cr} = 23$ MPa (at quenching depth in 1 K) the tendency of the system to polydomain ordering is completely suppressed (curves 3,4 in Fig. 5, 6(a) and 6(b)).

The study of initial stage of formation of domain structure (Fig. 5(b)) showed that compare to the potassium nitrate, the incubation period can be observed in KDP crystals (curves 1, 2 in Fig. 5(a)). The delay time while waiting for the nuclei of ferroelectric domains is fractions of a second. It was showed earlier that the duration of the incubation period is inversely proportional to the quenching parameter $t_{incub.} \sim 1/\alpha$ [12]. In our task the quenching parameter α depends on the value of pressure $p$ that was imposed on the sample. Therefore, the time delay of the formation of domain structure of KDP crystal under the hydrostatic pressure is defined as

$$t_{incub.} \sim \alpha^{-1} = \left(1 - \frac{p}{p_{cr}} - \frac{T}{T_C^{(0)}}\right)^{-1}, \quad p < p_{cr}. \tag{14}$$

As can be seen from expression (14), the higher is the pressure imposed on the system, the longer is the formation of the initial nuclei of ferroelectric domains. Thus, under the small pressure the relaxation starts almost immediately (curve 1 in Fig. 5(b)). The delay in the development of the domain structure increases with pressure (curves 2, 3 in Fig. 5(b)). It can be explained by "competition" between intrinsic elastic forces in the crystal and external hydrostatic pressure. Under a very strong pressure (upon condition $p > p_{cr}$) the expression (14) becomes negative and loses its physical meaning, i.e. the ferroelectric phase is destroyed.

If the pressure imposed on the sample is close to some critical pressure $p \sim p_{cr}$, both the formation of polydomain structure (curve 2 in Fig. 5(b)) and its destruction can happen (curve 3 in Fig. 5(b)). Under pressures $p \sim p_{cr}$ the nonequilibrium system is very sensitive to any external influences. Therefore, the final result of relaxation depends on the initial conditions. In this case the correction $T/T_C^{(0)}$ in (14) acquires a very important role, since the result of the evolution of the domain structure (development or destruction) is determined precisely by the quenching depth in temperature.

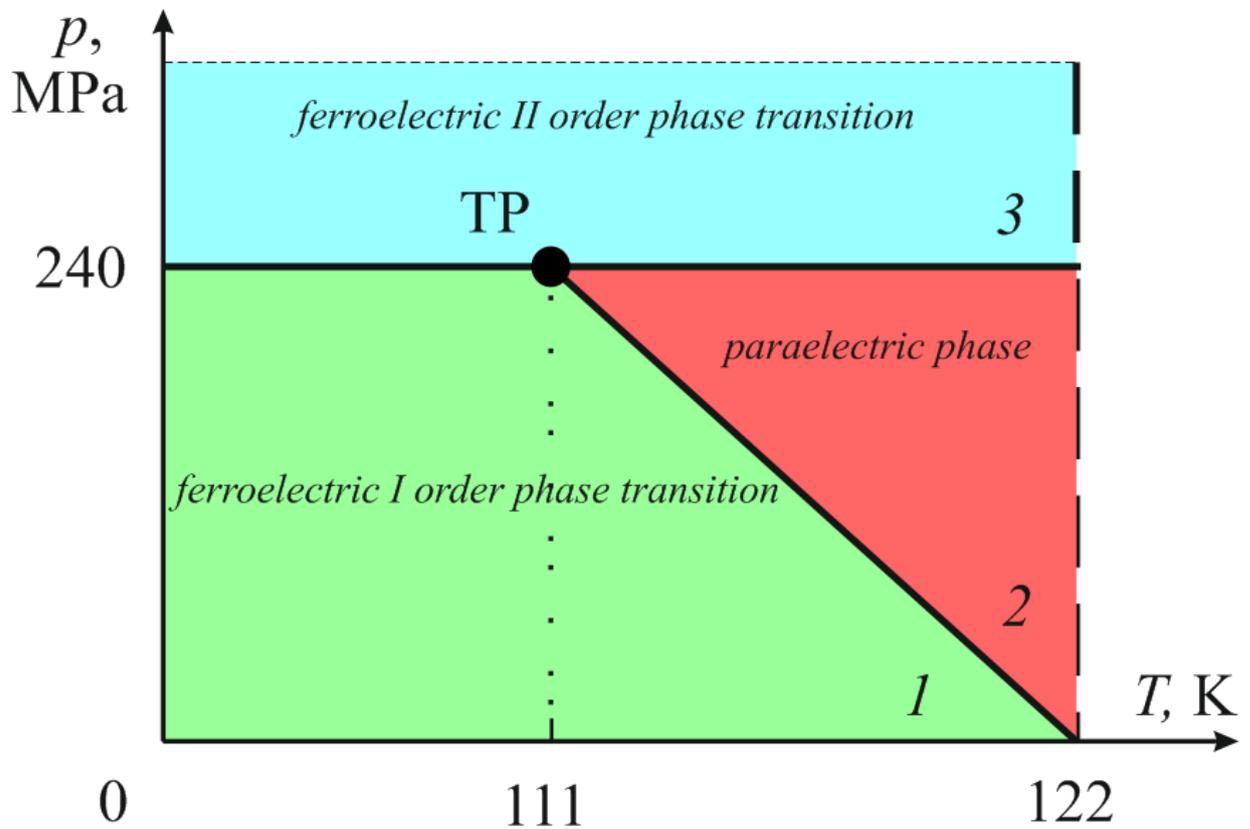

**Fig. 7.** Phase diagram $p – T$ with indication of the tricritical point TP. In sector *1*, a first-order ferroelectric phase transition takes place. Sector *2* defines the region of the disordered (paraelectric) phase under pressure. Sector *3* corresponds to second-order ferroelectric phase transitions

Numerical analysis confirmed that the critical pressure of domain structure destruction $p_{cr}$ significantly depends on quenching parameter α (Fig. 7). Thus, at pressures $p \leq 23$ MPa when quenching depth is 1 K ($T_1 = 121$ K) the system remains in the region of polydomain structure formation (sector 1 in Fig. 7). At pressures $p > 23$ MPa the system falls into the region of paraelectric phase (sector 2 in Fig. 7). The range of pressures that can be used to control the domain structure expands with increasing of quenching parameter α. At quenching depth in 11 K

($T_1 = 111$ K) the system cannot fall into sector 2, and the polydomain structure forms at any pressures (Fig. 7).

Numerical analysis showed that the tricritical point in phase diagram $p - T$ has coordinates $T = 111.1$ K and $p = 240$ MPa (Fig. 7). The tricritical point is defined as a point which separates the paraelectric phase and two ferroelectric phases (phase transitions of first and second order) [50]. The calculated value of tricritical point is in good consistent with known experimental data, where the TP point is defined as $p = 0.24$ GPa and $T \sim 111.5$ K [19]. At pressures more than tricritical value $p > 240$ MPa the system falls into sector 3 in phase diagram (Fig. 7) and the evolution of the domain structure of a ferroelectric proceeds according to scenarios characteristic of a second-order phase transition under pressure [18]. Thus, using the $p - T$ phase diagram, we can determine the values of the control parameters at which the formation of the domain structure will occur (Fig. 7). The developed theoretical model makes it possible to calculate the tricritical point for certain crystals. The calculations performed are in good agreement with the known data, which confirms the applicability and reliability of our model.

It is important to admit that the influence of hydrostatic pressure competes with the effect of electric field which is imposed on the sample at quenching stage $\varepsilon_q$. The electric field launches the mechanism of domain nucleation and forms some average polarization $(\bar{\pi}_0 \neq 0)$. After quenching the field turns off and the external pressure promotes rearrangement of the domain structure so as to minimize the internal energy of the crystal. Therefore, the higher the pressure is applied to the sample, the faster it eliminates the "displacement" effect of the quenching electric field $\varepsilon_q$, and the faster the relaxation of the system proceeds (curves 3, 4 in Fig 2; curves 1, 2 in Fig. 5). As a result, a stable polydomain structure with minimal average polarization $\bar{\pi}$ forms.

## 6. CONCLUSION

Kinetics of ordering of nonequilibrium system under hydrostatic pressure using ferroelectric crystals that undergo first order phase transitions of order-disorder type was studied. It was established earlier that these ferroelectrics tend predominantly to a monodomain ordering state in the absence of external influences and under external electric field [22]. However, the effect of hydrostatic pressure on the sample at relaxation stage provides a complex of domain structures.

Our analytical consideration is based on the previously applied model of the expansion of the Landau-Ginzburg-Devonshire functional in degrees of the order parameter [12]. It was shown that ferroelectrics, that undergo first order phase transition, are significant sensitive to the influence of hydrostatic pressure. Based on the statistical approach that we used previously [12, 18, 22] we obtained the system of nonlinear evolution equations for average polarization and its dispersion.

The qualitative analysis, made using phase pattern concept, allowed us to establish general regularities of ordering process of quenched nonequilibrium ferroelectric. It was shown that both monodomain and polydomain structures can form under the influence of hydrostatic pressure. The higher the value of the pressure imposed on the sample (in a range $p << p_{tcr} << \tilde{p}$) the more likely the probability of the formation of stable polydomain structure with high degree of inhomogeneity. It was shown that the relaxation process of the system can proceed both monotonically and with the formation of short-living metastable asymmetric polydomain phases.

Numerical analysis of obtained evolution equations was made for $KNO_3$ and KDP crystals which have different signs of baric coefficient.

Using $KNO_3$ crystal as an example it was shown that under hydrostatic pressure the system evolves predominantly to polydomain state of ordering. Monodomain structure can form only under the influence of a very weak pressure (or in the absence of it) at deep quenching, that corresponds to the results obtained previously in [22]. However, when the pressure is close to the tricritical value $p \rightarrow p_{tcr}$ and is imposed on the sample, the polydomain structure forms at any quenching depth. It can be explained by the fact that under such pressure the first order phase transition acquires features of second order phase transition for which the polydomain ordering state is more preferable [12, 18].

It was shown that relaxation time of domain structure significantly depends on the value of pressure. The higher the value of pressure $p$ the faster the system turns into a state of thermodynamic equilibrium. This dependence is especially noticeable with shallow quenching and under the small value of pressure.

Numerical analysis made using KDP crystal as an example showed that in this ferroelectric polydomain structure forms only, independently on quenching depth and value of pressure imposed on the sample. It was established that the incubation period – waiting time for nuclei of ferroelectric domains – can be observed at early relaxation stage. Duration of this time delay is inversely proportional to quenching depth and value of pressure.

It was established that in KDP crystals there is some critical pressure $p_{cr}$ at which the destruction of ferroelectric ordering occurs. The magnitude of this critical pressure $p_{cr}$ increases with quenching parameter α. As it turned out, under deep quenching ($T_C - T_1 > 10$ K), the ferroelectric phase will exist at pressures up to the tricritical value $p_{tcr}$.

Within the framework of analytical consideration general regularities of ordering process of ferroelectrics that undergo first order phase transition were obtained. However, the numerical analysis allowed us to establish the other characteristic features of ordering process using two ferroelectrics with different signs of baric coefficient γ as an example. Therefore, the results obtained can be used in practice for experiments with the application of certain ferroelectric crystals.

## ACKNOWLEDGMENTS

This work was supported by the Institute for Physics of Mining Processes of the NAS of Ukraine under the Grant No. 0117U002192.## DECLARATION OF INTERESTS

The authors declare that they have no known competing financial interests or personal relationships that could have appeared to influence the work reported in this paper.

## DATA AVAILABILITY

The data that support the findings of this study are available from the corresponding author upon reasonable request.

## REFERENCES

1. A. Chanthbouala, V. Garcia, R. Cherifi, K. Bouzehouane, S. Fusil, X. Moya, S. Xavier, H. Yamada, C. Deranlot, N. D. Mathur, M. Bibes, A. Barthélémy, J. Grollier, A ferroelectric memristor, Nature Materials 11 (2012) 860–864. https://doi.org/10.1016/j.Sc.2010.00372
2. T.Y. Kim, S.K. Kim, S. Kim, Application of ferroelectric materials for improving output power of energy harvesters, Nano Convergence 5 (2018) 30–46. https://doi.org/10.1186/s40580-018-0163-0